\documentclass[aps,pra,twocolumn,superscriptaddress,longbibliography]{revtex4-1}

\usepackage{amsmath}
\usepackage{graphicx}
\usepackage{epstopdf}
\usepackage{upgreek}
\usepackage[english]{babel}
\newcommand{\YIG}{\textsc{yig}}

\begin{document}

\title{Magneto-optical coupling in whispering gallery mode resonators}



\author{J. A. Haigh}
\affiliation{Hitachi Cambridge Laboratory, Cambridge, CB3 0HE, UK}
\author{S. Langenfeld}
\affiliation{Cavendish Laboratory, University of Cambridge, Cambridge, CB3 0HE, UK}
\author{N. J. Lambert}
\affiliation{Cavendish Laboratory, University of Cambridge, Cambridge, CB3 0HE, UK}
\author{J. J. Baumberg}
\affiliation{Cavendish Laboratory, University of Cambridge, Cambridge, CB3 0HE, UK}
\author{A. J. Ramsay}
\affiliation{Hitachi Cambridge Laboratory, Cambridge, CB3 0HE, UK}
\author{A. Nunnenkamp}
\affiliation{Cavendish Laboratory, University of Cambridge, Cambridge, CB3 0HE, UK}
\author{A. J. Ferguson}
\affiliation{Cavendish Laboratory, University of Cambridge, Cambridge, CB3 0HE, UK}


\date{\today}

\begin{abstract}
We demonstrate that yttrium iron garnet microspheres support optical whispering gallery modes similar to those in non-magnetic dielectric materials. 
The direction of the ferromagnetic moment tunes both the resonant frequency via the Voigt effect as well as the degree of polarization rotation via the Faraday effect. An understanding of the magneto-optical coupling in whispering gallery modes, where the propagation direction rotates with respect to the magnetization, is fundamental to the emerging field of cavity optomagnonics. 
\end{abstract}

\maketitle

\section{Introduction}

The high $Q$-factors and strong confinement in optical whispering gallery mode (WGM) resonators \cite{ilchenko_optical_2006,vahala_optical_2003} have enabled large coupling strengths to their collective modes of mechanical vibration. This has made these resonators central to the field of cavity optomechanics \cite{aspelmeyer_cavity_2014}, allowing parametric amplification and cooling of mechanical modes \cite{schliesser_radiation_2006} as well as near quantum limited measurement sensitivity \cite{schliesser_resolvedsideband_2009}.

To compliment optomechanics, it is worth considering alternative collective modes that could be coupled to the optical WGM resonators. The collective magnetization dynamics in ferromagnetic materials represents one such possibility. The magnetic field tunability of the these modes may have advantages over the fixed mechanical resonances, including easier adaptation to higher microwave frequencies. Recent work towards this has explored the strong coupling of microwave cavity photons to the magnetic modes in yttrium iron garnet (YIG) \cite{Huebl_high_2013, tabuchi_hybridizing_2014,zhang_strongly_2014,haigh_dispersive_2015,bai_spin_2015}. This interaction has enabled the coupling of single magnons to qubits \cite{tabuchi_coherent_2015} and multiple magnets together \cite{lambert_cavity_2015}. If this new system could be taken into the optical domain it would open further experiments comparable to those in optomechanics. In particular, the ease of coupling to both low frequency ($\lesssim$~GHz) and optical ($\gtrsim100$~THz) electromagnetic modes has led to the prospect of bidirectional quantum-coherent conversion between microwave and optical photons \cite{andrews_bidirectional_2014, bagci_optical_2014}.

In this Article, we demonstrate that the static direction of the magnetization can modulate the optical WGM frequencies in a similar way to mechanical position in optomechanics. This is achieved in YIG spheres exploiting well-known magneto-optical effects \cite{zvezdin_modern_1997,demokritov_microbrillouin_2008}, which can be large for the insulating YIG \cite{wettling_magnetooptics_1976}, and which are used in commercial devices such as optical isolators. The optical resonant frequencies are tuned via the Voigt effect, while the Faraday effect mixes the polarization of the linearly polarized modes. This shows that a magnet can be coupled to an optical cavity, opening up cavity optomagnonics to further experimental investigation.

\begin{figure}%
\includegraphics[width=0.85\columnwidth]{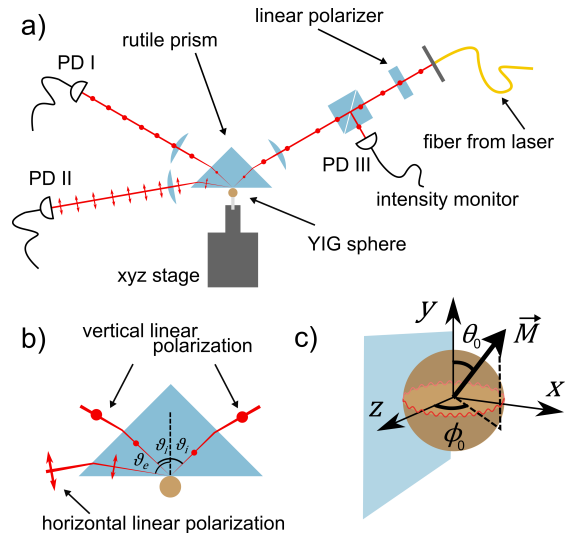}%
\caption{(a) Schematic of experimental setup. The light from the fiber-coupled laser is collimated, linearly polarized vertically (perpendicular to the plane), and focused onto the internal surface of the coupling prism. Two photodiodes on the output of the prism monitor the reflection of the input beam with vertical polarization (I) and the emission of any horizontally linear polarized light (II) from the YIG sphere. To account for fluctuations in laser intensity, all measurements are normalized to the input power monitored with a beamsplitter and photodiode (III) on the input path. The YIG sphere is mounted on a $xyz$ positioning stage to locate it at the coupling point. (b) The optical axis of the birefringent rutile prism is out-of-plane, which separates the output angles of the two linear polarizations for analysis via the two photodiodes. (c) Definition of the polar ($\phi_0$) and azimuthal ($\theta_0$) angles of the magnetization $\vec{M}$.}%
\label{diagram}%
\end{figure}

\section{Experimental setup}

A schematic of the experimental setup is shown in Fig.~\ref{diagram}. We use highly polished ferrimagnetic ($T_C\approx560$~K) YIG spheres of various radii, mounted on ceramic rods. An objective lens focuses the output of a narrow-linewidth ($\approx100$~MHz) external-cavity tunable diode laser onto the internal surface of the coupling prism \cite{gorodetsky_highq_1994}, with polarization set linearly along the vertical axis. The evanescent coupling to the YIG sphere is measured in the totally internally reflected beam on a photodiode (PD I). To match the wavevector of the incident light in the prism to the WGM the coupling angle of incidence $\vartheta_i$ is set to the critical angle for total internal reflection at the prism-sphere interface \cite{gorodetsky_optical_1999,schunk_identifying_2014}. The rutile coupling prism is birefringent with $n_e\approx2.7$ and $n_o\approx2.4$ (c.f.~$n_{\YIG}\approx2.2$ at 1300~nm). Due to this birefringence the two linear polarizations parallel and perpendicular to the plane of the WGM are out-coupled at different angles. Photodiode (PD II) measures the light emitted from the cavity with orthogonal polarization to the input beam, that is, linearly polarized in the incidence plane. Prism coupling is chosen over tapered \cite{knight_phasematched_1997} or polished \cite{serpenguzel_excitation_1995} fiber methods as the free space optics allows careful control of the input and measured light polarization as well as easier matching of the wavevector at the coupling point \cite{gorodetsky_optical_1999}.

\section{Observation of whispering gallery modes}

\begin{figure}%
\includegraphics[width=1\columnwidth]{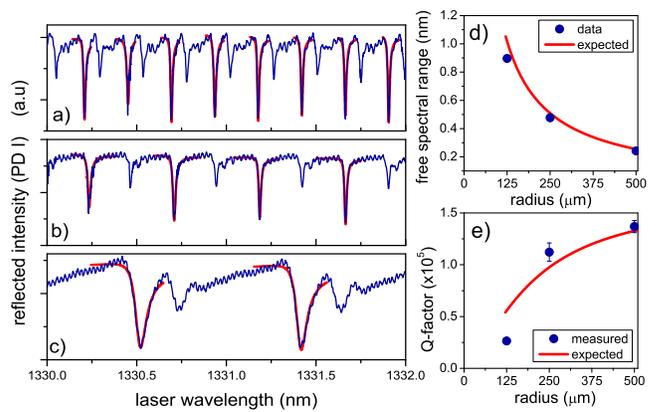}%
\caption{Reflected intensity (PD I) as a function of input laser free-space wavelength $\lambda$ for varying radius $r$ YIG spheres: (a) 500~$\upmu$m, (b) 250~$\upmu$m, and (c) 125~$\upmu$m. Red lines are separate fits to each peak, which give the peak separation (free spectral range) (d) and $Q$-factors (e). The red line in (d) is the expected free spectral range for the WGMs, $\Delta\lambda$, given by Eq.~(\ref{eq:FSR}), and the red line in (e) is the expected $Q$-factor.}%
\label{radii}%
\end{figure}

Figure~\ref{radii}~(a-c) show typical reflection spectra (PD I) for three YIG spheres of radii, $r=500,250,125~\upmu$m. A periodic structure of dips is observed as a function of free-space wavelength $\lambda$. As plotted in Fig.~\ref{radii}~(d), this period is inversely proportional to the radius of the sphere and is in quantitative agreement with the expected free spectral range for the whispering gallery modes given by
\begin{equation}
\Delta\lambda = \lambda^2 / (2 \pi r n_{\YIG}).
\label{eq:FSR}
\end{equation}
The good agreement identifies these resonances as the whispering gallery modes. The data show two families of peaks. The lower intensity dips are from the excitation of higher order WGMs due to an overlap with the coupling angle and position of the input beam. For the rest of the measurements we concentrate on the larger intensity dip, which we attribute to the lowest order WGM family.

To extract the $Q$-factor of the WGMs the dips are fitted individually with a Lorentzian lineshape, including an anti-symmetric component, present due to an angle-dependent phase shift of the reflected beam on the internal surface of the prism \cite{hecht_optics_2002}. From these fits we plot the $Q$-factors as a function of radius in Fig.~\ref{radii}~(e). The values are much smaller than those typical in non-magnetic WGM resonators, where extraordinarily high $Q$-factors of $\sim$10$^9$ are achievable \cite{collot_very_1993,vernooy_highq_1998}. This is to be expected, as, although YIG is transparent in the infrared, the absorption length is much shorter than in optical glasses. Using a reasonable value for the absorption coefficient $\alpha\approx0.1$~cm$^{-1}$ \cite{wood_effect_1967} to find the internal $Q$-factor $Q_i= {n_\YIG}/{\alpha\lambda}$ and the analytical expression for the loading $Q$-factor $Q_c$ \cite{gorodetsky_optical_1999},
\begin{equation}
Q_c =\frac{\pi}{2}\left(\frac{\omega r}{c}\right)^{\frac{3}{2}}(n_\YIG^2-1)\sqrt{\frac{n_\YIG}{n_e^2 - n_\YIG^2}},
\label{eq:qfactor}
\end{equation}
we can find the expected $Q$-factor $1/Q=1/Q_i+1/Q_c$. This is plotted as the red line in Fig.~\ref{radii}~(e) and is in reasonable agreement with the measured values, further confirming that the WGMs can be clearly identified. The dissipation rate ($\sim2$~GHz) is of the same order as the typical ferromagnetic resonance frequencies of YIG and, as the absorption coefficient in YIG is known to decrease considerably with temperature \cite{wood_effect_1967}, higher $Q$-factors can be expected in low temperature experiments.

\section{Magneto-optical properties}

We next outline the effect of the direction of the magnetization of the sphere on the WGMs. The Faraday and Voigt effects \cite{dillon_linear_1969}, corresponding to first and second order magnetization corrections to the dielectric tensor \cite{zvezdin_modern_1997}, make the WGMs substantially different to those observed in non-magnetic dielectric spheres. The Faraday effect depends linearly on the magnetization component in the direction of propagation of the light and results in a difference in the refractive index for the two circular polarizations, commonly observed as a rotation of the linear polarization in transmission through a ferromagnetic material. The Voigt effect is a birefringence quadratic in the magnetization, which results in a different refractive index for the two linear polarizations parallel and perpendicular to the magnetization direction. While in non-magnetic WGM resonators the modes are horizontally and vertically linear polarized, split by the difference in boundary conditions at the interface,  the off-diagonal terms in the dielectric tensor due to the Faraday effect mix these modes. In addition, the direction of propagation of the light relative to the magnetization changes with position around the WGM.

To analyze the effect of magneto-optical birefringence, we construct a simple model based only on the lowest order WGM. We consider a single channel with a length given by the circumference of the sphere. This ignores all radial and azimuthal mode structure, treating the propagation of light in the structure as a plane wave, but with a wavevector that varies around the mode. The displacement field $\vec{D}=(D_h,D_v)$, in the rotating wave approximation, satisfies $\boldsymbol{\upeta}(\vec{k})\vec{D} - (\omega^2/c^2)\vec{D}/k^2 = 0$, where $\boldsymbol{\upeta}$ is the inverse dielectric tensor in the plane perpendicular to the direction of propagation $\vec{k}$ \cite{landau_chapter_1984}, tangential to the sphere surface. In a medium with both linear and circular birefringence \cite{berry_adiabatic_1986}
\begin{equation}
\boldsymbol{\upeta} = 
\left(
\begin{array}{cc}
 \frac{1}{n_\YIG^2} +b_h^2+b_g & b_h b_v+i g_k \\
 {b_h} {b_v}-i {g_k} & \frac{1}{n_\YIG^2}+{b_v}^2-{b_g} \\
\end{array}
\right).
\label{eq:dielectric}
\end{equation}
The components of the linear $b_h,b_v$, and circular $g_k$, birefringence are given by ${b_h}=b \sin{\theta} \sin{\phi}$, ${b_v}=b \cos{\theta}$, and ${g_k}=g \sin{\theta} \cos{\phi}$ with $b$ and $g$ the magnitudes of the Voigt and Faraday effects. These components depend on the projection of the magnetization $\vec{M}$ onto a local coordinate system $(\hat{h},\hat{v},\hat{k})$, with orthogonal axes along the horiztonal, vertical, and in the direction of propagation. The angles $\theta$ and $\phi$ are those of the magnetization to the $\hat{v}$ and $\hat{k}$ axes, respectively, so that on the path around the mode $\theta=\theta_0$ is constant, while $\phi$ varies from $\phi_0$ to $\phi_0+2\pi$, with the angles $\theta_0$ and $\phi_0$ defined relative to the lab ($\hat{x},\hat{y},\hat{z}$) axes (see Fig.~\ref{diagram}~(c)). The boundary conditions in the WGM geometry result in an effective birefringence for linear polarization perpendicular and parallel to the sphere surface. This geometrical linear birefringence is taken into account via the constant $b_g$ which does not depend on the position around the sphere and is estimated by comparing with the mode splitting in non-magnetic WGM resonators \cite{Gorodetsky_geometrical_2006,schunk_identifying_2014},
\begin{equation}
b_g \approx  \frac{c}{\omega}\frac{1}{r} \frac{\sqrt{n_\YIG^2-1}}{n_{\YIG}^4}.
\label{eq:}
\end{equation}

The geometrical linear birefringence dominates the anisotropic parts of the inverse dielectric tensor, so that the eigenvalues can be approximated by the diagonal terms, and the eigenmodes remain predominantly horizontally or vertically linear polarized. We therefore label the modes $h$ and $v$. The wavevectors of the eigenmodes $k_{h,v}$ are found from the eigenvalues of $\boldsymbol{\upeta}$ (Eqn. \ref{eq:dielectric}),
\begin{equation}
\begin{split}
\upsilon_{h,v} = &\frac{1}{n_\YIG^2}+\frac{1}{2} \left(b_h^2+b_v^2\right) \\ 
&\pm  \sqrt{\left(\frac{1}{2}\left(b_h^2-b_v^2\right) + b_g\right)^2+ b_h^2 b_v^2 + g_k^2},
\end{split}
\label{eq:evalues}
\end{equation}
using $\upsilon_{h,v} = \frac{\omega^2}{c^2} \frac{1}{k_{(h,v)}^2}$. Neglecting the small terms, $g_k^2$ and $b_h^2 b_v^2$, we obtain
\begin{align}
k_h &= \frac{\omega n_\YIG}{c} \left(1  - \frac{n_\YIG^2}{2}b_g - \frac{n_\YIG^2}{2}b^2 \sin^2{\theta} \sin^2{\phi}\right),\\
k_v &= \frac{\omega n_\YIG}{c} \left(1  +  \frac{n_\YIG^2}{2}b_g - \frac{n_\YIG^2}{2}b^2 \cos^2{\theta}\right).
\label{eq:k_v}
\end{align}

\subsection{Resonant wavelengths}

\begin{figure}%
\includegraphics[width=1\columnwidth]{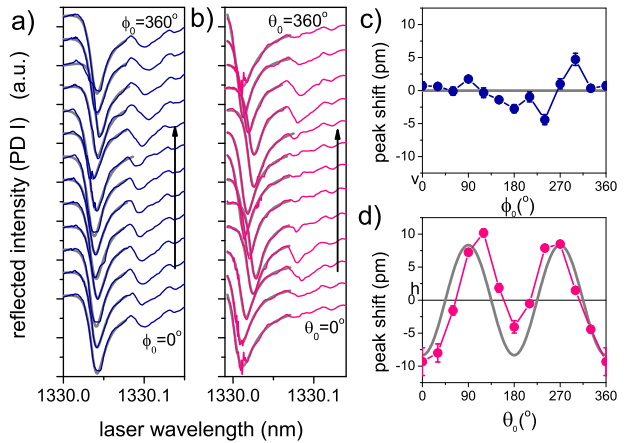}%
\caption{Reflected intensity (PD I) on a single resonance as a function of (a) in-plane angle $\phi_0$ ($\theta_0=90^{\circ}$) and (b) out-of-plane angle $\theta_0$ ($\phi_0=45^{\circ}$) of the magnetization with respect to the WGM plane. The fitted gray lines are used to extract the resonant frequency (c), (d), as a function of the two angles. The gray lines in (c) and (d) are the expected shifts considering only the Voigt effect given by Eq.~(\ref{eq:lambda_v}).}%
\label{field_v}%
\end{figure}

To find the dependence of the resonant wavelength on the magnetization direction, the wavevectors of the two eigenmodes are integrated over the path around the WGM to find the total phase accumulated for each mode. In principle, we should include the geometrical phase due to the change in basis around the WGM \cite{berry_adiabatic_1986,berry_budden_1990}.  However, this contribution is negligible as the solid angle subtended by the path on the Poincar\'e sphere is always small. From the resonance condition, $\int_{0}^{2\pi}{k_{h,v}rd\phi}=l\pi$ for integer $l$, the resonant wavelength can be found,
\begin{align}
\lambda_h &= \lambda_{h,0}\left(1 + \frac{n_\YIG^2}{4}b^2 \cos^2{\theta_0}\right)\label{eq:lambda_h},\\
\lambda_v &= \lambda_{v,0}\left(1 - \frac{n_\YIG^2}{2}b^2 \cos^2{\theta_0} \right),
\label{eq:lambda_v}
\end{align}
in terms of the resonant wavelengths $\lambda_{h,0}$ and $\lambda_{v,0}$ in the absence of magnetic birefringence. The resonant wavelength depends on the direction of the magnetization with respect to the vertical, $\theta_0$, and is determined only by the linear birefringence $b$ due to the Voigt effect. In contrast, the Faraday effect does not affect the optical mode frequency in a significant way, as it is suppressed by competition with the strong geometrical birefringence intrinsic to the WGM geometry.

To measure the magnetization direction dependence of the WGM frequencies, we use the stray field from a neodymium iron boron magnet to rotate the magnetization both in and out of the plane of the WGM. The magnitude of this magnetic field is $\approx80$~mT, enough to overcome the internal demagnetizing fields which act to break the sphere into ferromagnetic domains at low field \footnote{the field magnitude is obtained via the ferromagnetic resonance of the YIG sphere, measured through the change in reflection of a shorted co-axial cable positioned close to the sphere with a vector network analyzer}. Figure~\ref{field_v}~(a),(b) presents a series of reflection spectra as the magnetization is rotated (a) in-plane and (b) out-of-plane. The 500~$\upmu$m sphere is used, and we concentrate only on a single WGM resonance. As expected, for the in-plane rotation there is no systematic change in the spectrum as the magnetization is rotated. However, for the out-of-plane rotation there is clear shift in the resonant wavelength. The resonant wavelengths extracted from fitting the spectra are plotted in Fig.~\ref{field_v}~(c),(d). For the out-of-plane rotation, the resonant wavelength oscillates with the expected $\cos^2\theta_0$ dependence given by Eq.~(\ref{eq:lambda_v}). The amplitude of $\approx15$~pm is consistent with a Voigt coefficient $k_V\approx75^{\circ}\text{cm}^{-1}$, as reported in Ref.~\cite{castera_isolator_1977}. Some slight deviation of the expected values (gray lines) and measurements are due to fluctuations of the ambient temperature. These measurements show that in ferromagnetic WGM resonators the optical resonance frequency can be modulated via the direction of the magnetization, with a magnitude and symmetry consistent with the Voigt effect.

\subsection{Coupling of linearly polarized modes}

Next, we consider the consequences of the Faraday effect on the WGM properties. Due to the symmetry of the mode there is no net component of the magnetization along the optical path for any direction of the magnetic field, and it might be expected that the Faraday effect is not important. However, the polarization of the eigenmodes of propagation on the path are modified, and this can affect the coupling of the linear polarized input beam and the measurement basis to the modes in the sphere. This can be seen by looking at the eigenvectors of $\boldsymbol{\upeta}$ (Eq.~(\ref{eq:dielectric})), which have the form of Jones vectors giving the polarization of the eigenmodes. These are $\vec{\nu}_h = (1,-\beta^*)$, $\vec{\nu}_v = (\beta,1)$, where
\begin{align}
&\beta =\label{eq:beta1} \\
&\frac{b_h b_v - i g_k}{\frac{1}{2}(b_h^2 - b_v^2) + b_g + \sqrt{\left(\frac{1}{2}\left(b_h^2-b_v^2\right) + b_g\right)^2+ b_h^2 b_v^2 + g_k^2}}.
\nonumber
\end{align}
To first order in $g$, this reduces to
\begin{equation}
\beta =  \frac{b_h b_v - i g_k}{(b_h^2 - b_v^2) + 2 b_g}.
\label{eq:beta2}
\end{equation}
This parameter quantifies the mixing of the two modes, proportional to the Faraday term $g_k$ and the diagonal linear birefringence $b_h b_v$, normalized to the dominant total linear birefringence between the horizontal and vertical $(b_h^2 - b_v^2) +  2 b_g$. The modes have either predominantly horizontal or vertical linear polarization as $\beta\ll1$.

\begin{figure}%
\includegraphics[width=1\columnwidth]{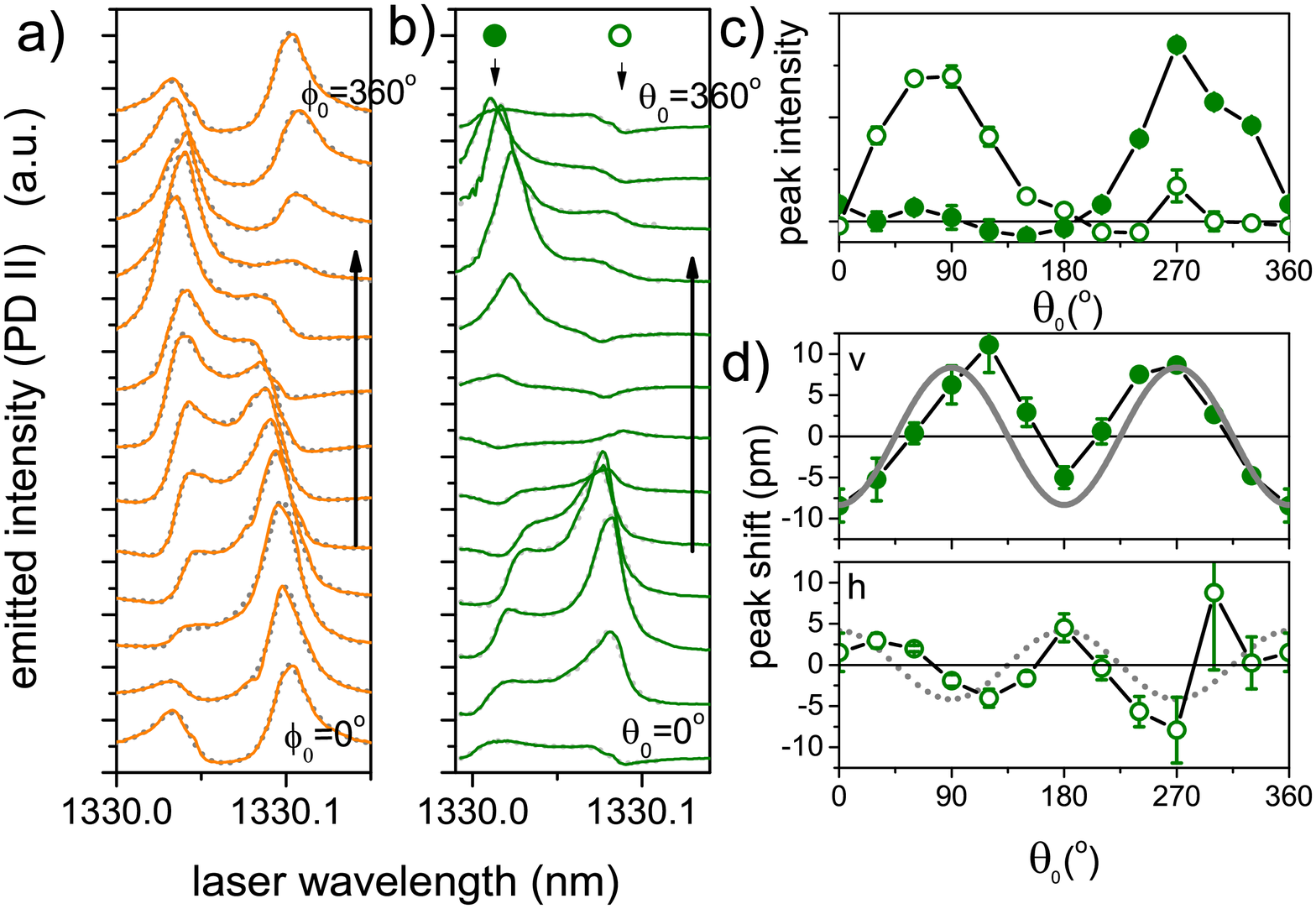}%
\caption{Emitted intensity from WGM with horizontal linear polarization (PD II), opposite to the input beam. Emitted intensity as a function of laser wavelength for (a) in-plane angle $\phi_0$ and (b) out-of-plane angle $\theta_0$ of the magnetization. Gray dotted lines are double peak fit to data. (c) Peak intensity of the two peaks as a function of out-of-plane angle $\theta_0$ (from fitting in (b)). (d) Wavelength shift of the two peaks as a function of out-of-plane angle $\theta_0$. The gray lines are the expected dependence based on Voigt effect, same as Fig.~\ref{field_v}, for the two modes with mainly vertical (solid) and horizontal (dotted) linear polarizations.}%
\label{field_h}%
\end{figure}

To study the coupling between the polarized modes, we measure the horizontally polarized light output from the YIG sphere with a second photodiode (PD II), as shown in the diagram in Fig.~\ref{diagram}~(a). For a non-magnetic WGM resonator we would not expect any emission into this orthogonal polarization. The spectra are shown in Fig.~\ref{field_h}~(a),(b). There are two peaks with relative amplitudes depending strongly on the direction of the magnetization both in- and out-of-plane. The lower wavelength peak corresponds to the main dip in the reflection spectra (Fig.~\ref{field_v}), and therefore is assigned to the mostly vertically polarized mode. Importantly, if the magnetization is out-of-plane there is no emitted intensity at either wavelength, as shown in the peak intensities plotted as a function of out-of-plane angle in Fig.~\ref{field_h}~(c). At these angles, $\theta_0=0^\circ$ and $180^\circ$, there is no Faraday effect because there is no component of the magnetization along the propagation direction of the light at any point around the WGM, and correspondingly $\beta=0$.

The two peaks are identified with what in the non-magnetic limit are the vertical and horizontal linear polarized modes of the sphere. They have amplitudes that depend on the mixing of the two linear polarizations at the coupling point. Based on our simple model, we expect the two peaks to have the same amplitude. However, because of a small amount of background light received at PD II, which interferes with the emitted light, the lineshapes depend on their different relative phases, and it is difficult to compare the actual intensities.

Further evidence of the polarization mixing is observed as a dip in the magnetic field dependent reflection spectra shown in Fig.~\ref{field_v}~(b) at the same wavelength as the higher wavelength peak in emission. The amplitude of this dip depends on the direction of the magnetization, and is only present when there is an in-plane component, consistent with the Faraday effect.

The mixing of the linear polarized modes provides access to the wavelength shifts of the predominantly horizontally linearly polarized mode, which would not otherwise be observed with vertically polarized input beam. The frequency shifts of the two modes measured in emission are plotted in Fig.~\ref{field_h}~(d), as a function of out-of-plane angle $\theta_0$. The lower wavelength mode matches with that measured in reflection (Fig.~\ref{field_v}~(d)), and with Eq.~(\ref{eq:lambda_v}). The higher wavelength mode also changes with angle with the opposite sign and a smaller amplitude, as given by Eq.~(\ref{eq:lambda_h}). This further confirms the validity of assigning these shifts largely to the Voigt effect.

\section{Modeling of spectra}

We can use our model to make direct comparison with the measured spectra. The transmission matrix of the cavity is calculated by taking the phases accumulated by the eigenmodes in propagation around the path and transforming back into the linear polarization basis that we measure in using the eigenmodes at the coupling point. The diagonal matrix of the phase factors $\mathbf{A}$ is rotated by the column matrix of eigenvectors, $\mathbf{V} = (\vec{\nu_+},\vec{\nu_-})$,  $\mathbf{T} = \mathbf{V} \mathbf{A} \mathbf{V}^{-1}$.  We obtain
\begin{align}
\mathbf{T}& = e^{i \Gamma_0 }\times \\
&\left(
\begin{array}{cc}
 \cos{\Gamma_d} - i \sin{\Gamma_d} \frac{\beta  \beta ^* - 1}{\beta  \beta ^* + 1} &  2 i \sin{\Gamma_d} \frac{ \beta^*  }{\beta  \beta ^*+1} \\
 2 i \sin{\Gamma_d} \frac{ \beta  }{\beta  \beta ^*+1} & \cos{\Gamma_d}+ i \sin{\Gamma_d} \frac{ \beta  \beta ^* -1 }{\beta  \beta ^* +1 } \\
\end{array}
\right).\nonumber
\end{align}
This can be put into the matrix equation 
\begin{align}
\vec{D}_{\text{out}} =& (1-\kappa_c)\mathbf{R}\vec{D}_{\text{in}} +  {\kappa_c} [ (1- {\kappa_i}) \mathbf{T} - 1]^{-1} \vec{D}_{\text{in}}\label{eq:matrixEqn1}\\
=& \mathbf{S} \vec{D}_{\text{in}}\nonumber
\end{align}
to find the output field from the prism-cavity system, in terms of scattering matrix $\mathbf{S}$. Here $\mathbf{R}$ takes into account the phase shift in the totally internal reflected light (see appendix), and the $\kappa$, $\kappa_{\text{c}}$ are the single-cavity-pass losses for internal dissipation and external coupling respectively. These are found from the corresponding $Q$-factors, $Q_i$ and $Q_c$,
\begin{equation}
\kappa_{i,c} = 1 - e^{-{2 \pi r n_\YIG}/{\lambda Q_{i,c}}}.
\label{eq:kappa}
\end{equation}
A small amount ($\sim0.5\%$) of experimentally observed scattered light is added to the horizontal channel which reproduces the measured asymmetry between the two peaks in emission.

\begin{figure}%
\includegraphics[width=\columnwidth]{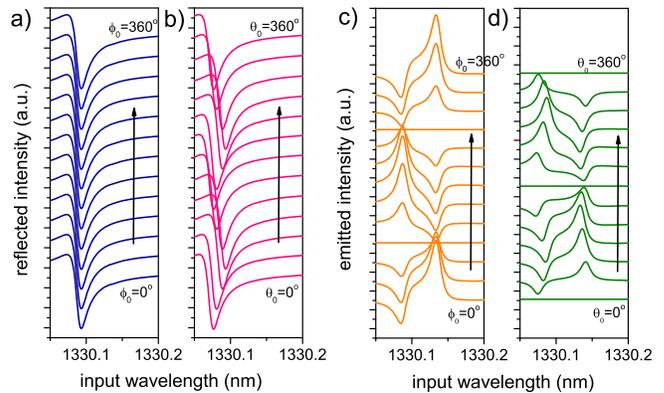}%
\caption{Calculated transmission spectra ($|S_{22}|^2$) for rotation of the magnetization (a) in-plane ($\phi_0=0\to360^\circ$, $\theta_0=0$) and (b) out-of-plane ($\theta_0=0\to360^\circ$, $\phi_0=45^\circ$). These correspond to the measured data in Fig.\,2 (a,b). Calculated emission spectra ($|S_{21}|^2$) for rotation of the magnetization (c) in-plane ($\phi_0=0\to360^\circ$, $\theta_0=0$) and (d) out-f-plane ($\theta_0=0\to360^\circ$, $\phi_0=45^\circ$). These correspond to the measured data in Fig.\,3 (a,b).}%
\label{fig5}%
\end{figure}

The results of this calculation are plotted in Fig.\,\ref{fig5}. We see that this simple model quantitatively reproduces the main features of the data. This justifies the asymmetric Lorentzian fit, and neglecting of higher order corrections to the accumulated phase and eigenmode polarizations.

\section{Conclusions}

In this Article we have characterized the interaction of the optical WGM and the magnetization in the static limit. This provides a natural starting point for analyzing the dynamical properties of the system. The parametric coupling of the magnetization to the optical mode through the Voigt effect may enable analogous experiments to those performed in cavity optomechanics \cite{aspelmeyer_cavity_2014}. There the coupling is understood in terms of the changes in the path length of the cavity, typically linearly modulated by the mechanical position. In contrast, our measurements show that the largest coupling is quadratic in the magnetization. In addition, the Faraday effect couples the two ordinarily linear polarized modes. The small mode volume of the optical cavity may enhance magnon Brillouin scattering amplitudes, towards the magnetic analog to Raman phonon lasing in non-magnetic WGM resonators \cite{spillane_ultralowthreshold_2002}. The magnetic field tunablilty of the ferromagnetic resonance modes may have advantages over the static mechanical modes of opto-mechanics, allowing complimentary experiments and applications.

During the final stages of preparation of this manuscript we became aware of related works \cite{osada_cavity_2015,zhang_optomagnonic_2015}.

\section*{Acknowledgments}
We are grateful for useful discussions with Mete Atature and Richard Phillips. This project was partly funded by EPSRC under EP/M50693X/1. A.~J.~F. is supported by ERC grant 648613 and a Hitachi research fellowship. A.~N. holds a University Research Fellowship from the Royal Society and acknowledges support from the Winton Program for the Physics of Sustainability.

\appendix*

\section{Phase shift on internal reflection}

The non-symmetric Lorentzian lineshape of the measured resonances are due to phase shifts $\varphi_{h,v}$ of the reflected beam on the internal surface of the prism. These can be calculated from the Fresnel equations \cite{hecht_optics_2002},
\begin{align}
\tan{\varphi_h} =& -\frac{2 \cos\vartheta_i \sqrt{n_o^2 \sin^2{\vartheta_i} - 1}}{(n_o^2 - n_o^4) + 2 (1 + n_o^4)\cos^2\vartheta_i},
\label{eq:varphi_h}\\
\tan{\varphi_v} =& -\frac{2 \cos\vartheta_i \sqrt{n_e^2 \sin^2{\vartheta_i} - 1}}{(1 - n_e^2) + 2 n_e^2\cos^2\vartheta_i}.
\label{eq:varphi_v}
\end{align}
We include these in the matrix equation (\ref{eq:matrixEqn1}) through the matrix
\begin{equation}
\mathbf{R}=
\left(
\begin{array}{cc}
e^{-i\varphi_h} & 0\\
0 &  e^{-i\varphi_v}
\end{array}
\right).
\label{eq:R}
\end{equation}

\bibliographystyle{apsrev4-1}
\bibliography{bibfile}

\end{document}